\documentstyle[12pt]{article}
\begin{document} 
\begin{flushright}{OITS 700}\\
March 2001
\end{flushright}

\vspace*{1cm}

\begin{center} 
{\Large {\bf Void Analysis of Hadronic Density Fluctuations
at Phase Transition}}
\vskip .75cm
 {\bf  Rudolph C. Hwa$^1$ and Qing-hui Zhang$^2$}
\vskip.5cm
 
\vskip.5cm
 
{$^1$Institute of Theoretical Science and Department of
Physics\\ University of Oregon, Eugene, OR 97403-5203, USA\\
\bigskip
$^2$Physics Department, McGill University, Montreal QC 
H3A 2T8, Canada}
\end{center}
\vskip.5cm

\begin{abstract} 
The event-to-event fluctuations of hadron multiplicities are
studied for a quark system undergoing second-order phase
transition to hadrons.  Emphasis is placed on the search for
an observable signature that is realistic for heavy-ion
collisions.  It is suggested that in the 2-dimensional
$y$-$\phi$ space the produced particles selected in a very
narrow $p_T$ window may exhibit clustering patterns even
when integrated over the entire emission time.  Using the
Ising model to simulate the critical phenomenon and taking
into account a $p_T$ distribution that depends on the
emission time, we study in the framework of the void
analysis proposed earlier and find scaling behavior.  The
scaling exponents turn out to be larger than the ones found
before for pure configurations without mixing.  The
signature is robust in that it is insensitive to the precise
scheme of simulating time evolution.  Thus it should reveal
whether or not the dense matter created in heavy-ion
collisions is a quark-gluon plasma before hadronization.
\end{abstract}
\section{Introduction}

It is well known that a system undergoing a second-order
phase transition exhibits large fluctuations and long-range
correlations.  If the symmetry of a quark system created in
a heavy-ion collision is such that the quark-hadron phase
transition is second order, then one would expect large
fluctuations in the multiplicity of hadrons produced, not
only from event to event (which happens normally in
high-energy collisions), but also from one region to another
in the geometrical space into which the particles are
emitted at any given time.  It is the latter type of
fluctuations that we focus on as a possible signature of the
phase transition, when an appropriate measure can be
identified.

The fluctuation of local hadron density implies the
formation of spatial patterns that exhibit clusters of
hadrons of varying sizes.  There are also regions of no
particles, which are referred to as voids.  An investigation
of the fluctuation of void sizes was carried out by us
recently in Ref. \cite{hz}.  We found scaling behaviors,
whose exponents were proposed as an observable signature
of quark-hadron phase transition (PT).

Recently an analysis of the NA49 data on $Pb$-$Pb$
collisions at CERN-SPS was carried out according to the
method described in Ref. \cite{hz}.  It is found
\cite{cy} that the scaling behaviors exist under certain restricted
conditions of the analysis, and that the scaling exponents are
significantly lower than the ones predicted in \cite{hz}.

The void analysis studied in \cite{hz} has its limitations.
The most serious shortcoming is that it is an analysis of a
system that is essentially static.  The aim has been to find a
measure suitable for quantifying effectively a given spatial
pattern, or a set of them.  However, the hadronization
process in a heavy-ion collision is dynamical and takes
place over an extended period of time.  If the particles
emitted throughout the whole duration are collected at
every local region, then the patterns examined in the static
scenario overlap one another, and the fluctuating patterns
would disappear in the cumulative result.

An analogy to this problem is the nature of rainfalls in two
geographical regions with very different climates.  It is not
possible to judge from their annual precipitations
(assuming that they are roughly the same) whether the
rain in one region is stormy, or the other drizzly.  It is
necessary to make cuts in the time duration to learn about
the spatial fluctuations in short time intervals.  Similarly, in
heavy-ion collisions it would be ideal if we could select a
small $\Delta t$  interval to analyze the spatial pattern of
an event.  That is, however, not feasible in a heavy-ion
experiment.  The only cut that can be made experimentally
while keeping the $y$-$\phi$ space for analysis is in the
$p_T$ variable, i.e., selecting particles in a small $\Delta
p_T$ interval.  How to prepare the spatial configuration for
our analysis in a way that corresponds to how the
experimental data are to be analyzed is an essential part of
the present work.

As in \cite{hz}, which follows the theoretical framework or
Ref. \cite{hwu}, we assume that the dense quark matter
created in a heavy-ion collision is cylindrical in shape,
expanding rapidly in the longitudinal direction and more
gradually in the radial direction.  It is further reasonable to
assume that the thermal system is hot on the inside (with
$T > T_c$) and cooler on the surface, where at $T \simeq
T_c$ the plasma undergoes a PT to hadrons.  Thus the
critical phenomenon occurs on a two-dimensional surface
with the particle momenta being mainly normal to that
surface.  The fluctuation that we want to examine is in the
hadron density from region to region in the $y$-$\phi$
space for particles that are emitted in a small $p_T$
interval.  To simulate that fluctuation we make use of the
theoretical result that the phase boundary between a
first-order PT for some low quark masses and a crossover
at higher quark masses is a second-order PT belonging to
the same universality class as the Ising system
\cite{ggp,kr}.  We map a section of the $y$-$\phi$ space
on the surface of the plasma cylinder created in a heavy-ion
collision to a 2D Ising lattice.  We use the Ising model to
simulate the spatial pattern of clusters of spins pointing in
the same direction, and relate a cell with a net spin
pointing  in one direction to a hadron, and a cell with a net
spin in the other direction to no-hadrons, i.e., locally
and temporarily still in the quark state.  That is the model
that we have used in
\cite{hz,hwu}, the essential part of which will be
summarized in Sec.\ 2.

Regarding what is simulated on the Ising lattice as a
representation of the hadrons density fluctuation on the
surface of the plasma cylinder at any given time, we have to
relate that to the experimental situation of particles
collected in a $\Delta p_T$ cut,  but accumulated
throughout the entire hadronization time that can exceed
10 fm/c.  It is a problem of preparing the
temporally-integrated configuration on the lattice that can
best approximate the real data.  The problem will be
discussed in Sec.\ 3.  The analysis of the mixed
configurations that leads to scaling behaviors is the content
of Sec.\ 4.

\section{Simulation and Analysis of Hadronic Density\\
Fluctuation}

An essential feature of critical behavior is that the critical
exponents of different systems belong to the same
universality class if the systems have the same symmetry
and dimension \cite{ng}.  Our use of the Ising model to
simulate the spatial patterns at PT is based on the
recognition that for appropriate masses of the light and
strange quarks the quark-hadron PT is in the same
universality class as the Ising system \cite{ggp,kr}.  That is
fortunate, since the simulation of Ising lattice is extremely
simple.  However, the realistic quark masses may not have
exactly the values that correspond to the ones on the phase
boundary where the second-order PT occurs.  It is highly
likely that the realistic point in the phase diagram is in the
broad region of a crossover above the boundary line
characterized by the Ising system.  Fortunately again, a
crossover can also be easily simulated on the Ising lattice
by providing a preferred direction for the spins to be
aligned.  We summarize here the main points in the
relationship between the Ising problem and the hadron
density problem, and then outline the void analysis, the
details of which can be found in \cite{hz}.

For a 2D lattice of size $L^2$ with each site having spin
$\sigma_j = \pm 1$, we define the spin aligned along the
overall magnetization $m_L = \sum_{j \in L^2}\sigma_j$
by
\begin{eqnarray}
s _j = \mbox{sgn} (m_L) \sigma _j
\quad ,
\label{1}
\end{eqnarray}
where $\mbox{sgn} (m_L)$ stands for the sign of $m_L$.
We relate hadronization at location $i$ to a cell of size
$\epsilon^2$ with net spin $c_i$ positive, where
\begin{eqnarray}
c_i =
\sum_j s_j \quad ,
\label{1.5}
\end{eqnarray}
with the sum being over all sites in the
cell.  If $c_i \leq 0$, no hadronization takes place at $i$.
Hadron density is defined by
\begin{eqnarray}
\rho_i =  \lambda c^2_i \theta (c_i)\quad ,
\label{2}
\end{eqnarray}
where $\lambda$ is an unspecified factor relating the
lattice spins to the number of particles per unit area in
$y$-$\phi$ space.  The scaling behavior to be examined
should, of course, be independent of the normalization
factor $\lambda$.

Since lattice spins fluctuate from site to site when $T$ is
near $T_c$, $\rho_i$ also fluctuates from cell to cell, either
positive or zero.  When averaged over all cells of a
configuration, and then averaged over all uncorrelated
configurations at any given $T$, $\left<\rho\right>$ is a
smooth function of $T$, descending rapidly near $T_c$ as
$T$ is increased \cite{hz, hwu}.  It is a crossover, since the
first few orders of derivatives exhibit no signs of
discontinuities.  The value of $T_c$ is determined by
examining the scaling behavior of the normalized factorial
moments $F_q$ \cite{zc}, and is found to be $T_c = 2.315$
in the Ising model units of $J/k_B$.

In practice we work with a lattice having $L = 288$ and
cells of size $\epsilon = 4$.  For scaling behavior we study
the dependences of our measures on the bin size $\delta
\times \delta$.  The average hadron density in a bin is
$\bar{\rho _b} = \nu^{-1}\sum^{\nu}_{i = 1} \rho _i$,
where $\nu = (\delta/\epsilon)^2$.  Near $T_c$,
$\bar{\rho _b}$ fluctuates from bin to bin, especially for
small $\delta$.  It is that fluctuation we want to
characterize by quantitative measures.  Our proposal is to
represent $\bar{\rho _b}$ in the 2D space by topographical
maps with various discrete elevations $\rho_0$.  For every
given $\rho_0$ we define a bin to be ``empty'' when
\begin{eqnarray}
\bar{\rho _b}< \rho_0  \quad .
\label{3}
\end{eqnarray}
and further define a void to be a collection of contiguous
empty bins, connected by at least one side between
neighboring bins.  An analogy for this analysis is to flood a
rough terrain by water and characterize the landscape
profile by either land (nonempty bin) or water (empty
bin).   Clearly, by studying the areas of the submerged
regions at different water levels, one can capture the
topographical properties of the terrain.  In our problem the
maximum possible value of $\rho _i$ in a cell is
$(\epsilon^2)^2 = 256$ in units of $\lambda$.  A range of
$\rho _0$ between 20 and 100 have been used in our void
analysis in \cite{hz}.

For the void analysis let $V_k$ be the size of the {\it k}th
void in units of bins
\begin{eqnarray}
V_k = \sum_{\left< b\right>_k} \theta (\rho_0 -
\bar{\rho}_b)
\quad ,
\label{4}
\end{eqnarray}
where the sum is over all empty bins in the {\it k}th void.
Then, let $x_k = V_k/M$ be the fraction of bins in the
lattice occupied by $V_k$, where $M = (L/\delta)^2$ is the
total number of bins, and let $m$ be the total number of
voids in the configuration.
We define the normalized $G$
moments to be
\begin{eqnarray}
G_q = {1 \over m} \sum^m_{k = 1} x^q_k \ \left / \ ({1 \over
m}
\sum^m_{k = 1} x_k)^q \quad.\right.
\label{5}
\end{eqnarray}
For every given configuration, $G_q$ can be determined for
various bin sizes, i.e., as a function of $M$.  $G_q$ can
fluctuate widely from configuration to configuration.  That
fluctuation can be described by a probability distribution
$P(G_q)$.  While many moments of that distribution can be
studied, we have proposed the two lowest moments
\begin{eqnarray}
\left<G_q\right> = \int dG_q\,G_q\,P(G_q) \quad ,
\label{6}
\end{eqnarray}
\begin{eqnarray}
S_q = \left<G_q\, {\rm ln}\, G_q\right> \quad ,
\label{7}
\end{eqnarray}
the latter being the  derivative with respect to $p$ of
the {\it p}th moment evaluated at $p = 1$ \cite{hz}.

For pure configurations simulated on the Ising lattice,
scaling behaviors have been found for both
$\left<G_q\right>$ and
$S_q$
\begin{eqnarray}
\left<G_q\right> \propto M^{\gamma_q} \quad ,
\label{8}
\end{eqnarray}
\begin{eqnarray}
S_q \propto M^{\sigma_q} \quad ,
\label{9}
\end{eqnarray}
for a range of values of $\rho_0$ \cite{hz}.  Furthermore,
$\gamma_q$ and $\sigma_q$ depend on $q$ linearly as
\begin{eqnarray}
\gamma_q = c_0 + cq
\quad ,
\label{10}
\end{eqnarray}
\begin{eqnarray}
\sigma_q = s_0 + sq
 \quad .
\label{11}
\end{eqnarray}
Thus the values $c$ and $s$ (which are the slopes of
slopes) are concise characterizations of the fluctuation
behavior near the critical point.

It should be stressed that although the 2D Ising model can
be solved exactly for an infinite system\cite{kh}, the
conventional description of its critical behavior is in terms
of the temperature $T$ in the vicinity of $T_c$.   For a
quark-hadron system created in a heavy-ion collision, $T$
is not measurable, nor controllable.  Our proposed
measures of the critical behavior can be determined in
high-energy nuclear experiments without any knowledge
about $T$.

\section{Mixed Configurations for an Evolving System}

In the previous section we described the observable
measures that can be applied to the heavy-ion data on the
one hand, and can extract salient features of critical
configurations on the other.  Now we focus on the issue of
preparing the configurations that correspond more closely
to the realistic situation of a quark-gluon plasma expanding,
cooling, and hadronizing.

While a detailed modeling of the hydrodynamical
expansion of a quark-gluon system is inappropriate for our
purpose here, we can emphasize the main features of such
an expanding system and incorporate them in our
preparation procedure.  Our assumption is that the
temperature profile of the plasma cylinder is such that $T$
is high $(> T_c)$ in the interior and decreases
monotonically toward the surface until $T_c$, where
quark-hadron PT takes place at the surface.  The hadrons
are emitted nonuniformly from the surface with transverse
momenta $p_T$ that can be measured for each particle.
The average $\left<p_T\right>$ depends on the radial
pressure gradient, which in turn depends on time, since the
longitudinal expansion lowers the temperature throughout
the cylindrical interior and the radial expansion lowers the
radial pressure gradient.  Disregarding high-$p_T$ hard
scattering, which affects $\left<p_T\right>$ only in a
minor way, we may take the general hydrodynamical
implication to be that $\left<p_T\right>$ decreases with
the pressure gradient, and therefore with the evolution
time $t$.

If the $p_T$ distribution at any given time were narrow,
but when integrated over all times it gives the usual
exponential damping in $p_T$ for the event distribution
$dN/dp_T$, then our problem would be simple.  In the
idealized situation of a one-to-one correspondence between
$p_T$ and $t$, a narrow $\Delta p_T$ cut in $p_T$ would
result in a narrow $\Delta t$ cut in emission time.  That
would provide us with a narrow temporal window to view
the spatial pattern of hadronization activities on the
cylinder surface, and the pure configurations simulated on
the Ising lattice would suffice to model the effects of the
critical behavior, as discussed in the previous section.

The reality is, however, quite the opposite.  At any given
$t$ the $p_T$ distribution is broad, so the particles
emitted into a small $\Delta p_T$ interval can come from a
wide range of $t$.  That introduces a complication related
to the correlation between successive emission times.  But,
first, let us describe the $p_T$-$t$ correlation by a model
formula that captures the essence of the hydrodynamical
character of the $p_T$ distribution at different times
\cite{ph}.  We adopt the Gaussian formula
\begin{eqnarray}
P(p_T, t) = N(t) \mbox{exp}\{-[p_T - p_0(t)
]^2/2p^2_0(t)
\} \quad ,
\label{12}
\end{eqnarray}
where
\begin{eqnarray}
p_0(t) = 2/t \quad {\rm GeV/c}
\label{13}
\end{eqnarray}
and $N(t)$ is the normalization factor that preserves the
condition
\begin{eqnarray}
\int dp_T P(p_T, t) = 1
\label{14}
\end{eqnarray}
for all $t \geq 1$.  It is clear that, as $t$ increases, the
Gaussian peak shifts toward lower $p_T$ and its width also
decreases correspondingly.  If we make a narrow $\Delta
p_T$ cut, say $0.3 < p_T < 0.35$ GeV/c, there are
contributions from a wide range of $t$ that are not
negligible.

We note that the accuracy of Eq. (\ref{12}) is not important
for our purpose, since firstly the $p_T$ distribution depends
on the evolution model and on many parameters in the
initial condition, and secondly we use it only for simulating
configurations at different times in the evolution process in
the framework of the Ising model, which is a simple
generator of PT configurations that ignore the realistic
complications of surface irregularity, quark density
fluctuations, surface tension, etc.  However, Eq. (\ref{12})
does contain the essence of overlapping $p_T$
distributions at successive times, a feature that we want to
incorporate in our simulation.  The variable $t$ in Eqs.\
(\ref{12}) and (\ref{13}) is used only in the sense of
representing time in the simulation (to be detailed below),
not as real time in heavy-ion collisions in some physical
unit.  We emphasize that the correspondences between the
lattice variables and realistic quantities in heavy-ion
experiments have never been fixed.  In Eq.\ (\ref{2}) we did
not specify $\lambda$ that relates lattice spins to hadron
density; the scaling behaviors we seek are independent of
$\lambda$.  Here we similarly do not know how to specify
the precise relationship between the number of steps in the
evolution of the Ising lattice and the actual time of
hydrodynamical evolution of a quark system.  We now
consider how many steps to take.

If at any given $t$ we have on the Ising lattice a
configuration of hadrons and voids, then it is very
reasonable to expect that at 1 fm/c later in real time the
previous void-region will have a higher probability to
hadronize than the region where hadronization has already
occurred.  Of course, there are fluctuations in local
temperature, in surface curvature, pressure gradient, quark
density, and other local properties of the statistical and
collective system, all of which are unknown and can affect
hadronization locally.  Our procedure is primarily to let the
Ising system evolve by updating the spin configuration
using the Wolff logarithm \cite{uw}.  Let $S$ denote the
number of steps of updating the Ising spin orientations that
corresponds to 1 fm/c in real time.  Not knowing what $S$
is, but realizing that the Ising configuration does not change
drastically in one or two steps, we adopt the following two
options.

\noindent {\it Option 1}.	$S = 5$

After every five steps of updating the Ising configuration,
we reverse the direction of what is ``up'' to allow the cells
in the void region in the previous configuration to
be preferred for hadronization.  That is, denoting the
number of Ising steps by $nS$, we let $t = n$ in Eq.\
(\ref{12}), and replace Eq.\ (\ref{1}) by
\begin{eqnarray}
s_j=(-1)^n sgn(m_L)\sigma_j .
\label{15}
\end{eqnarray}
Note that if $S$ were $1$, and the Ising configuration
changes very slowly, then Eq.\ (\ref{15}) applied to
(\ref{1.5}) and (\ref{2}) would suggest that $\rho_i>0$ and
$\rho_i<0$ regions approximately alternate as $n$
increases.  That is an extreme scenario.  By letting $S=5$, we
allow statistical readjustment of the lattice spins as time
progresses.  The $(-1)^n$ factor in (\ref{15}) is put in by
hand to take into account the hydrodynamical feature that
any quark region is likely to hadronize at a later time due to
radial expansion and cooling.

\noindent {\it Option 2}.	 $S = 50$

After fifty steps it is likely that some of void regions will
have evolved to contain $\rho_i>0$ cells.  It is then not
necessary to impose sign flip by hand, so we can continue to
use Eq.\ (\ref{1}) for $s_j$ and rely on the fluctuations of
the Ising spins to generate new configuration that
correspond to successive emission time.  That is, with the
number of Ising steps being $nS$, we use $t = n$ in Eq.\
(\ref{12}), and without the $(-1)^n$ factor in (\ref{15}) we
proceed with the evolution process.  This is an option
representing a scheme quite different from the one above.

We now describe the algorithm for generating the
temporally-integrated configurations (TIC) in either option
of identifying $t$.  At each $t$ we simulate on the Ising
lattice a pattern of hadrons and voids.  Let us call that a
frozen configuration.  At the location of every cell where
$\rho_i>0$, we assign a value of $p_T$ according to the
distribution $P(p_T,t)$ given in Eq.\ (\ref{12}).  We choose
to keep only those particles whose $p_T$ fall within the
interval
$\Delta p_T$ that we select, say $0.3<p_T<0.35$ GeV.  The
cell location of each of those selected particles are recorded
in a new configuration, which is to become the TIC.  Now let
$t$ increase by one step, repeat the above procedure, and
enter more particles into the TIC at the cell positions where
the selected hadrons are located.  We continue to do this up
to
$t_m$ steps, when the accumulated particles at various places
in the 2D space constitute our final TIC for that sum.  The
value of $t_m$ is determined by the requirement that the
average hadron density for the whole lattice is
approximately equal to that in the frozen case (no $t$
evolution and any $p_T$) at the same temperature $T$.  In
this way we generate a TIC, starting always with $t = 2$ so
that $p_0 (t)$ ranges from $1$ down to 0.1 GeV/c.

For our analysis in the next section, we use a lattice of size
$L = 288$, with cells of size $\epsilon = 4$, and bin sizes
varying from $\delta = 8$ to $32$.  We select $\Delta  p_T$
to be between $0.2$ to $0.3$ GeV/c, and the corresponding
$t_m$ turns out to be $9$.  We generate $5 \times 10^3$
configuration for each sample for analysis, using both
options for $t$ evolution.

\section{Scaling Behaviors of the Temporally-Integrated\\
Configurations}

Having described how the TIC are prepared, we proceed
directly to the analysis discussed in Sec. $2$ and present the
results here.

Although the simulation can be done at any $T$ in the
vicinity of $T_c$ and the analysis can be done at any
elevation $\rho_0$, we show only the scaling behaviors for
$T=T_c$ and $\rho_0=40$ [in units of $\lambda$ in Eq.\
(\ref{2})].  We shall consider both options of $t$ evolution
discussed in the previous section and label the
corresponding figures by either $S=5$ or $S=50$.

In Fig.\ 1(a) we show $\left<G_q\right>$ vs $M$ in log-log
plot for
$S = 5$.  Clearly, it exhibits very good scaling behavior for
all values of $q$ considered.  It means that voids of all sizes
exist in the TIC.  Since the behavior is very similar to the
one found in \cite{hz} for the pure or frozen configurations,
the implication is that the mixing of configurations
selected in the $\Delta p_T$ window and integrated over
the emission time does not lead to a homogenization of the
clustering patterns.  This behavior is possible only if each
of the underlying (frozen) configuration exhibits critical
fluctuation.  Taking a small portion of each of such
configurations at different $t$ and adding them up retain the
fluctuating nature of the void structure that gives rise to the
scaling behavior.  This is a very fortunate feature that
makes possible the detection of the critical behavior in
heavy-ion collision.

From the slopes of the straight-line fits in Fig.\ 1(a), which
are denoted by $\gamma_q$ in Eq.\ (\ref{8}), we show the
dependence of $\gamma_q$ on $q$ in Fig. $1(b)$.  Evidently,
Eq.\ (\ref{10}) is a good description of that linear
dependence.  The slope $c$ of slopes is a numerical
characterization of the scaling behavior of
$\left<G_q\right>$.  Its value in this case is $c = 0.92$,
which is significantly larger than the value of $0.7$ found
in
\cite{hz} for pure (frozen) configurations.

The same analysis is carried out in option 2 of $t$
evolution.  The corresponding results are shown in Figs.\
2(a) and (b).  The value of $c = 0.97$ is slightly larger,
but is sufficiently close to the value from option 1 that
one can reasonably infer the essential independence of the
critical fluctuation on the procedure of simulating the
temporal evolution.

We have also investigated the dependences on $\rho_0$.
In Fig.\ 3 we show the values of $c$ for the two options in
$t$ evolution as functions of $\rho_0$.  It should be
understood that by giving the values of $c$ we are
asserting the existence of scaling such as those shown in
Figs.\ 1(a) and 2(a) for all the values of $\rho_0$
examined.  No essential difference between the two options
of time evolution has been found.  The feature in Fig.\ 3 to be
noted is the wide range of $\rho_0$ from 40 to 100 in which the
scaling behavior can be found.  To give a sense of the
normalization for those numbers, it should be remarked that
the average  $\left<\rho\right>$ is around 78.  That feature and
the values of $c$ ranging between 0.75 and 0.96 may now be
regarded as the quantitative signature of quark-hadron phase
transition that should be checked in heavy-ion collisions.

For all that have been done for $\left<G_q\right>$ above,
we can do the same for $S_q$ defined in Eq.\ (\ref{7}).
Fig.\ 4(a) shows the scaling behavior of $S_q$ vs $M$,
whose exponents $\sigma_q$ are shown in Fig.\ 4(b), all
for $S = 5$.  The value of the slope $s$, defined in Eq.\
(\ref{11}), is 0.82 and $\rho_0 = 40$.  For  $S = 50$ the
corresponding figures are shown in Figs.\ 5(a) and (b), and
the value of $s$ is 0.89.  Finally, the dependence of $s$ on
$\rho_0$ is shown in Fig.\ 6.  The range of $\rho_0$ is the same as
for $\left< G_q\right>$, and the range of $s$ is between 0.7 and 0.9.

For a rough comparison we mention that in the real data of
NA 49 the scaling behaviors of $\left<G_q\right>$ and
$S_q$ are found only in very restricted ranges of $\rho_0$
and the values of $c$ and $s$ are significantly lower
\cite{cy}.

For comparison, we have also performed the void analysis on
randomly generated statistical configurations. We have chosen
two methods to generate such configurations.

\noindent$\it Method\ 1.$\quad  We use random-number generator to
assign a value of $\rho_i$ between 0 and 160 at site
$i$, for $i$ ranging over all sites on the lattice. We choose that
range because  the average value of $\rho_i$ simulated in the
foregoing at $T_c$ is around 78. We collect $5\times 10^3$
such configurations and call them the random sample.

\noindent$\it Method\ 2.$\quad Generate first a configuration on
the $L\times L$ sites of the lattice in the Ising model as in
[1]. Choose randomly a cell $i$ and transfer the hadron density
$\rho_i$ there to a $\it new$ configuration at the same cell
location $i$. Repeat this procedure $(L/\epsilon)^2$ times, each
time starting with a new uncorrelated configuration;
furthermore, each time  the transfer of $\rho_j$ is made to
the $j$th cell of the same $\it new$ configuration. At the end of
the $(L/\epsilon)^2$th transfer, we have completed the
construction of a $\it new$ configuration, which is regarded as a
mixed configuration. Then we start afresh and generate a new
mixed configuration,  adding to what we call the statistical sample.
Specifically, we use $L=288$ and $\epsilon=4$. There are
no correlations among the various cells in these configurations,
although the underlying dynamics that generates the original
configurations is that of the Ising model.

The results of our void analysis  on the random and
statistical samples are very similar. Scaling behavior is not
found until $\rho_0$ is nearly as large as the mean density
and then only for a narrow range of $\rho_0$ between 60 and 80.
This is very different from our results shown in Figs.\ 3 and
6 that have a wide range of $\rho_0$ in which the scaling
behavior exists, beginning with $\rho_0\approx
\left<\rho\right>/2$.

Based on the above results we believe that dynamically
generated critical behavior in the experimental data can be
distinguished from the statistical background.

\section{Conclusion}

What we have demonstrated in this work is that critical
fluctuation in quark-hadron phase transition can be
observable in heavy-ion collisions, when appropriate
measures are sought for in the data analysis.  The details
of how we have simulated the configurations are not as
important as the insistence that a narrow $\Delta p_T$ cut
be made.  There is no chance that the experimental data at
RHIC can reveal any evidence of the pattern formation in
hadron density if such a cut in $p_T$ is not applied.
Scaling behavior is the canonical characteristics of critical
phenomena, so to discover its existence in the data of
heavy-ion collisions should be a primary area of
investigation for the signatures of quark-gluon plasma.
Since temperature is not directly measurable, and
fluctuations in multiplicities do not manifest themselves
naturally as in critical opalescence, our task is nontrivial and
must rely on the identification of observables that are
carefully constructed to reveal the scaling properties.  Of
course, if scaling is not found in the data, it could be that the
phase transition is of first order, or is a crossover of such
smoothness that no interesting fluctuations exist.  The
application of our method of analysis to the data can at
least shed some light on the nature of the phase transition.

There is a possibility that scaling behavior may exist even
in the absence of a second-order PT of the type discussed
here.  Such a possibility has been proposed earlier
\cite{hlp} as a manifestation of self-organized criticality
\cite{btw}.  Although the scaling discussed there is in
terms of cluster sizes, there is no doubt that a void
analysis can detect its signals, if it exists.  The method
proposed here should be sensitive to any hadronization
process that does not produce particles uniformly in a
short time interval.  Application of the method to the
RHIC data should be revealing whether or not scaling
behavior can be found.

\newpage

\section*{Acknowledgment}
One of us (RCH) wishes to acknowledge Dr.\ Pasi Huovinen
for his willingness to perform certain hydrodynamical
calculations and to transmit the results to us before
publication.  We are grateful to Prof.\ C.\ B.\ Yang for
communicating to us the preliminary result of his analysis
of the NA49 data.  This work was supported, in part,  by the U.\ S.\
Department of Energy under Grant No. DE-FG03-96ER40972,
the Natural Science and Engineering Research Council of Canada
and the Fonds FCAR of the Quebec Government.

\newpage
\begin{center}
\section*{Figure Captions}
\end{center}
\begin{description}

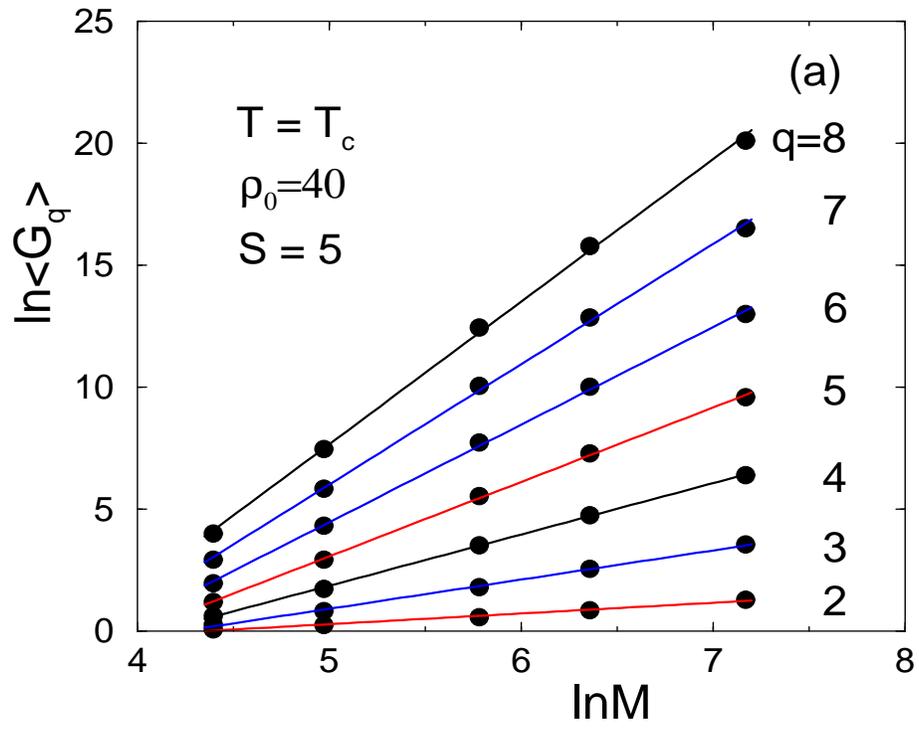
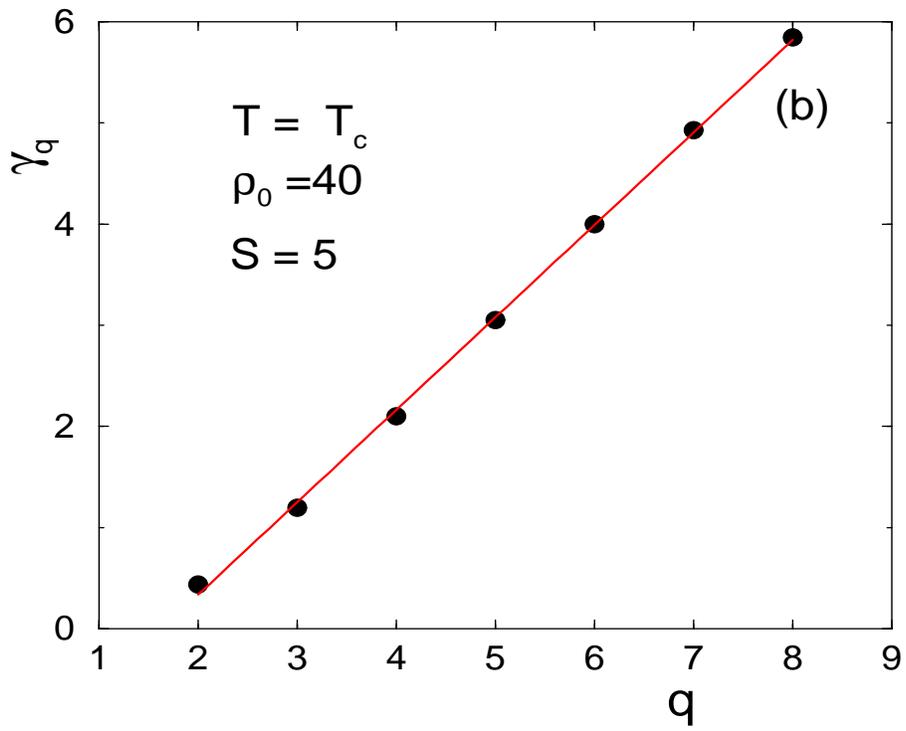
\item[Fig.\ 1]\quad Option 1 in time-evolution, $S=5$: (a)
Scaling behavior of $\left<G_q\right>$, (b) Dependence of the
scaling exponent $\gamma_q$ on $q$.

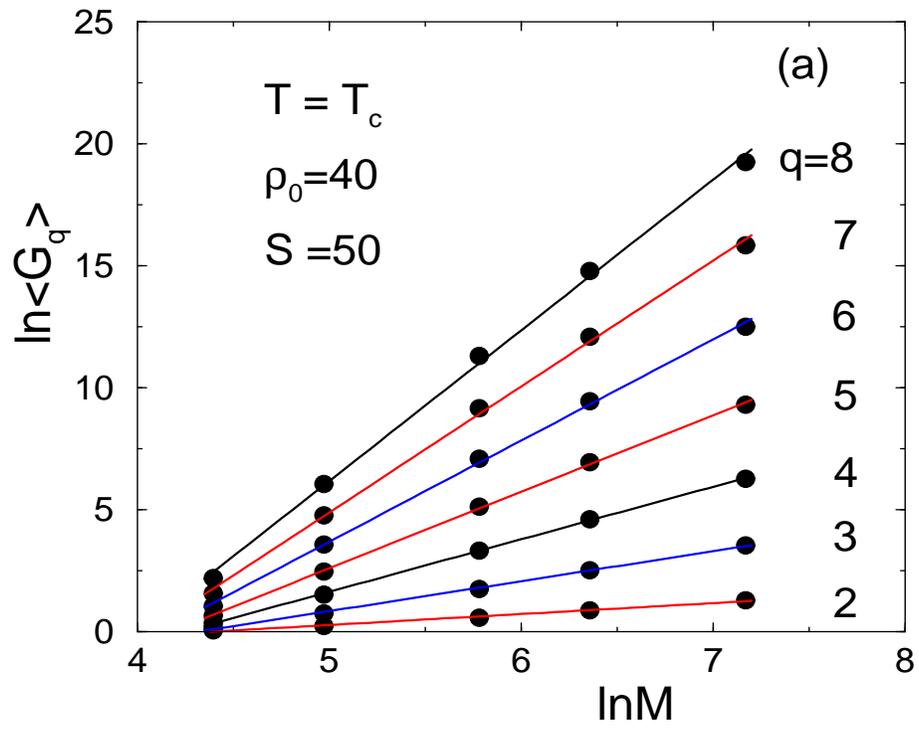
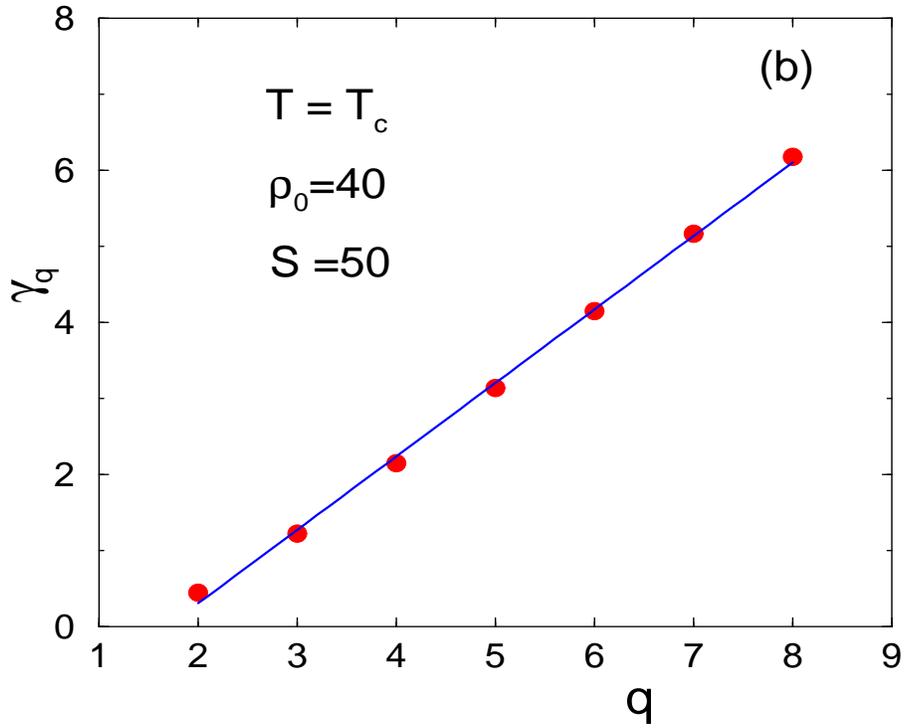
\item[Fig.\ 2]\quad Option 2 in time-evolution, $S=50$: (a)
Scaling behavior of $\left<G_q\right>$, (b) Dependence of the
scaling exponent $\gamma_q$ on $q$.

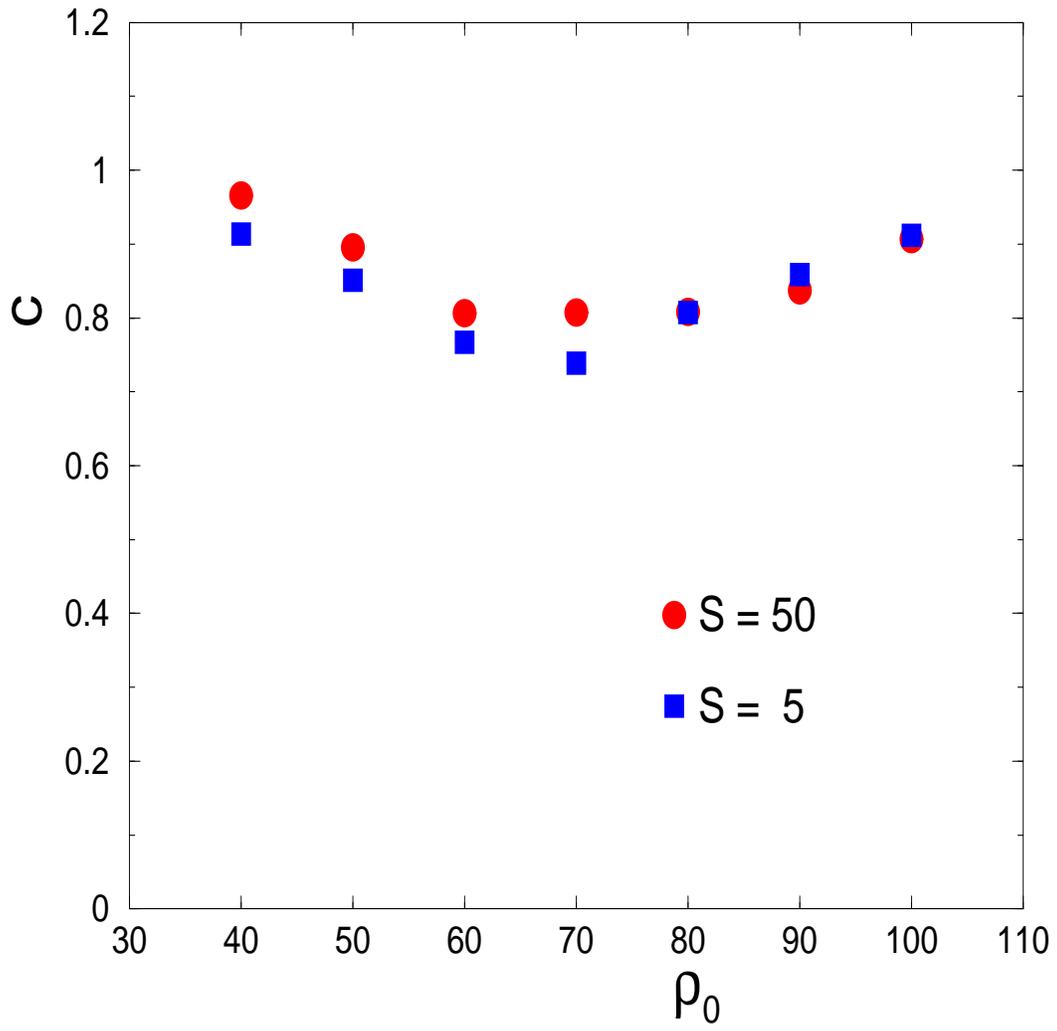
\item[Fig.\ 3]\quad Dependences of $c$ on $\rho_0$ for the
two options.

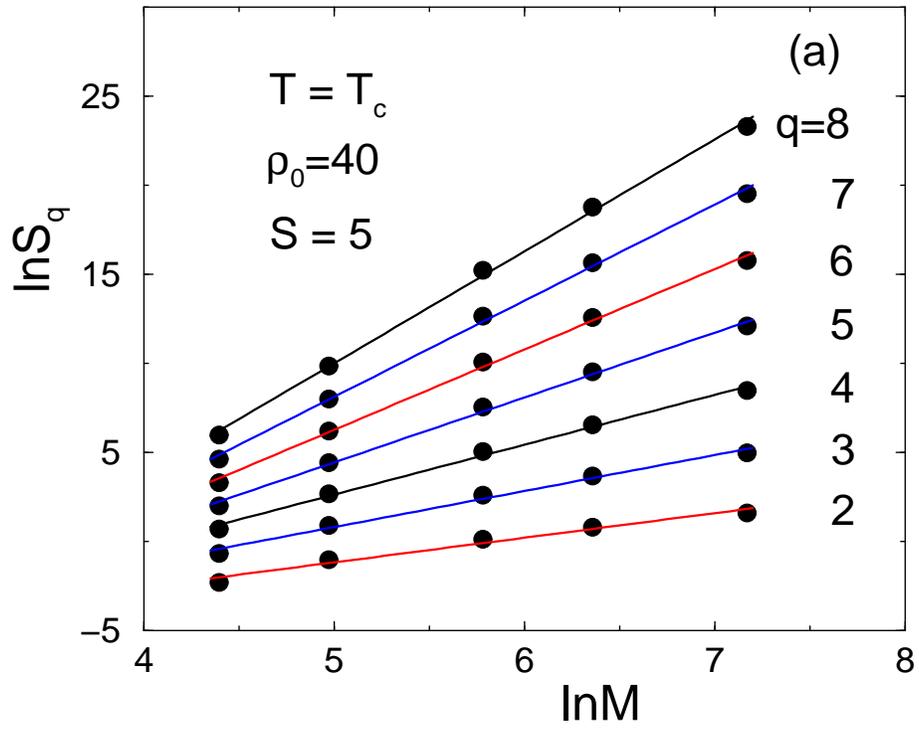
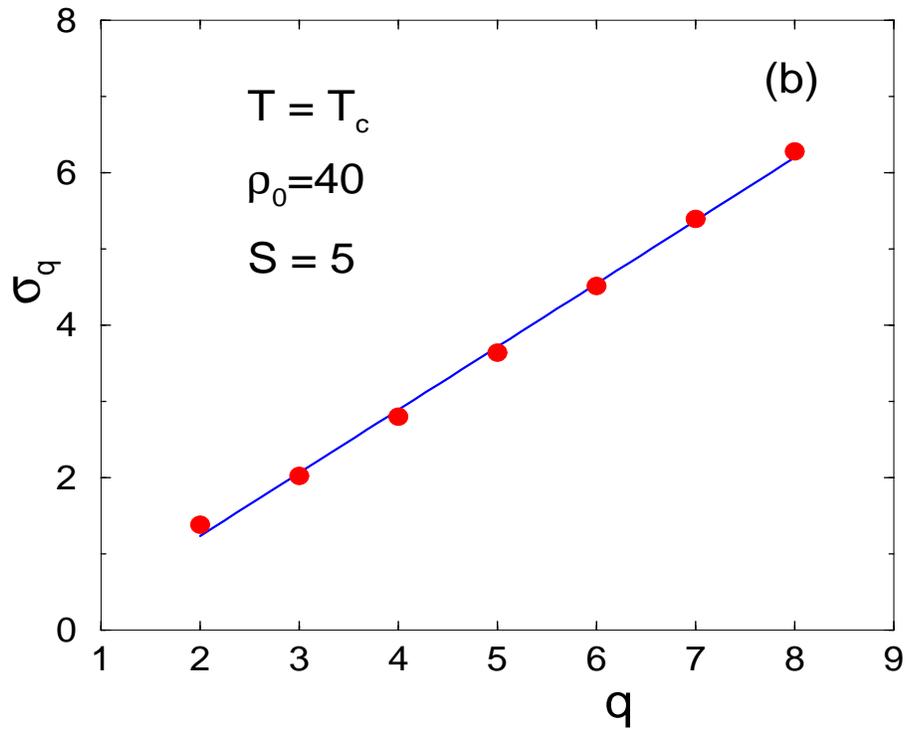
\item[Fig.\ 4]\quad Option 1 in time-evolution, $S=5$: (a)
Scaling behavior of $S_q$, (b) Dependence of the
scaling exponent $\sigma_q$ on $q$.

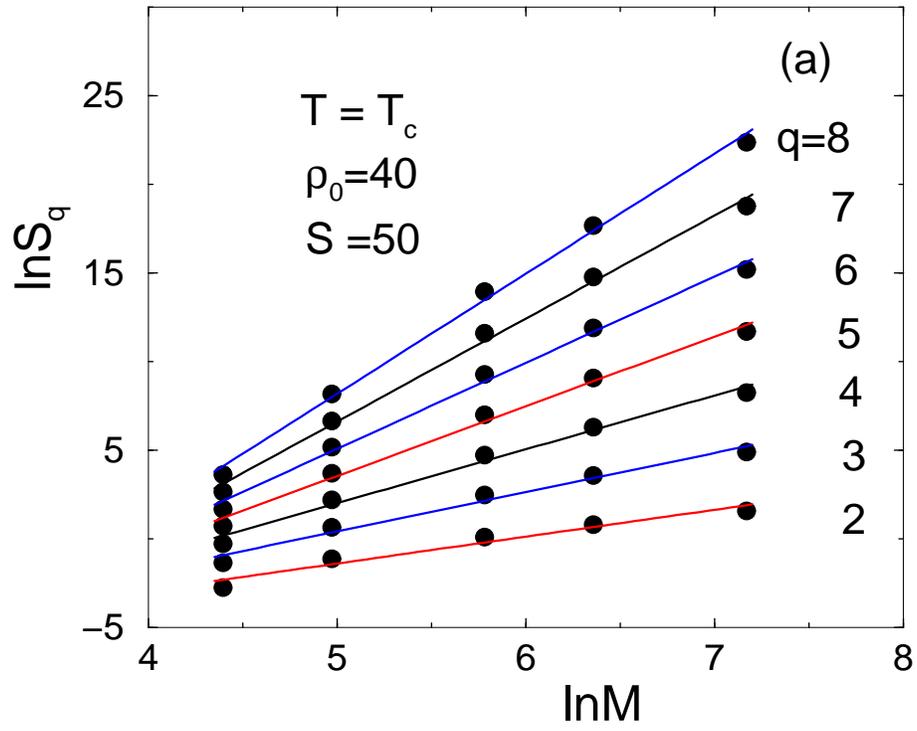
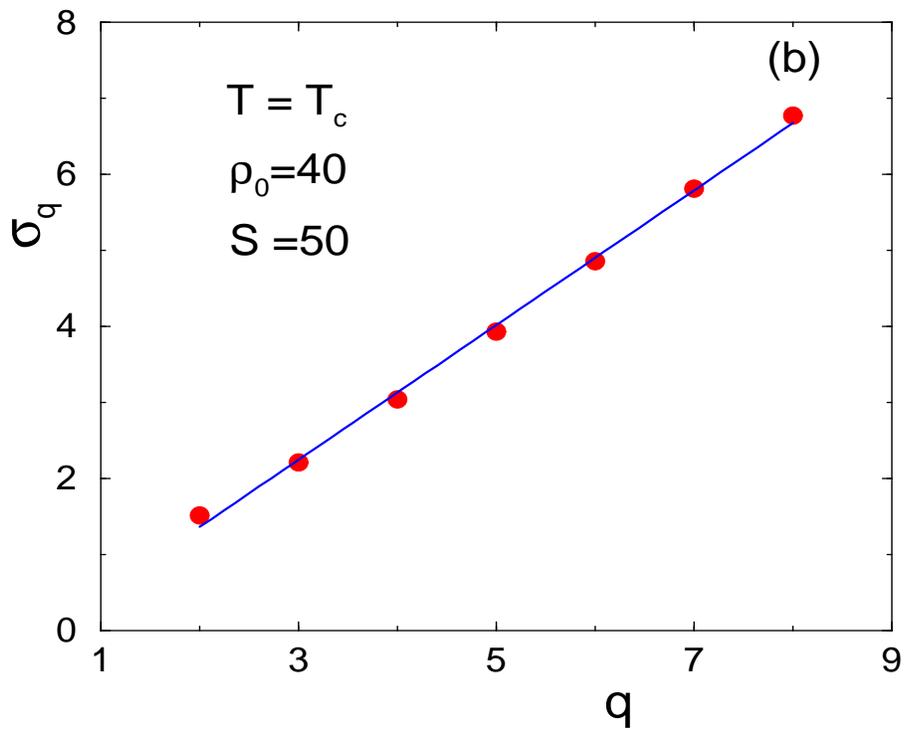
\item[Fig.\ 5]\quad Option 2 in time-evolution, $S=50$: (a)
Scaling behavior of $S_q$, (b) Dependence of the
scaling exponent $\sigma_q$ on $q$.

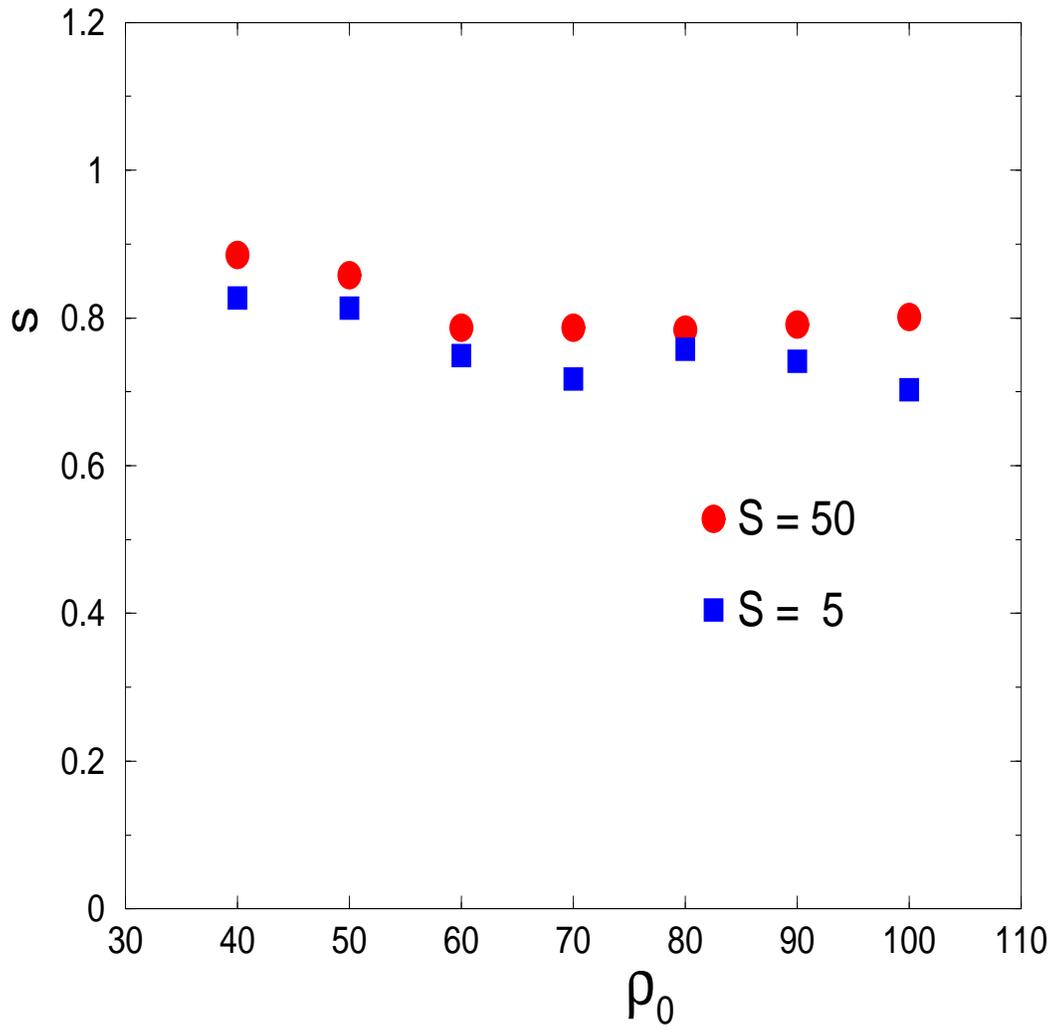
\item[Fig.\ 6]\quad Dependences of $s$ on $\rho_0$ for the
two options.

\end{description}

\end{document}